\documentstyle[12pt]{article}
\advance\textheight by 60pt
\advance\voffset by -40pt
\advance\textwidth by 50pt
\advance\oddsidemargin by -25pt
\advance\evensidemargin by -25pt

\def\single_space{\baselineskip 12pt plus 1pt minus 1pt}
\def\one_and_a_half_space{\baselineskip 19pt plus 1pt minus 1pt}
\def\double_spacesp{\baselineskip 25pt plus 2pt minus 2pt}

\def\atversim#1#2{\lower0.7ex\vbox{\baselineskip\zatskip\lineskip\zatskip
  \lineskiplimit 0pt\ialign{$\matth#1\hfil##\hfil$\crcr#2\crcr\sim\crcr}}}

\begin{document}
\begin{titlepage}
\begin{flushright}
{\bf
PSU/TH/184\\
June 1997\\
}
\end{flushright}
\vskip 1.5cm
{\Large
{\bf
\begin{center}
Closing the window on the axigluon mass \\
\vskip 0.01cm using top quark production data
\end{center}
}}
\vskip 1.0cm
\begin{center}
M.~A.~Doncheski \\
Department of Physics \\
The Pennsylvania State University\\
Mont Alto, PA 17237  USA \\
and \\
R.~W.~Robinett \\
Department of Physics\\
The Pennsylvania State University\\
University Park, PA 16802 USA\\
\end{center}
\vskip 1.0cm
\begin{abstract}

The contribution of axigluons (the massive color-octet gauge bosons in
all chiral color models) to top quark pair production in hadronic 
collisions is considered. The agreement between the experimental values
of the $t\overline{t}$ production cross-section at the TEVATRON and recent
QCD predictions is used to discuss limits on the axigluon mass.  Specifically,
intermediate mass axigluons, those in the mass range 
$50\,GeV < M_A < 120\,GeV$ which has not already been excluded, 
would increase the tree-level 
$q\overline{q} \rightarrow t\overline{t}$ cross-section by a factor of 
$\geq 2$, thereby increasing the theoretical predictions 
for $\sigma_{t\overline{t}}$ by $\Delta \sigma_{t\overline{t}} 
= 3.2-3.7\,pb$ ($2.7-3.1\,pb$) using leading-order 
(next-to-leading order) parton distributions over this mass range,
independent of the axigluon decay width. Such an increase is 
roughly $1.3-1.6$ ($0.9-1.2$) standard
deviations larger than that suggested by the apparent good agreement between
combined experimental results and recent theoretical calculations and so
is not ruled out, but is definitely disfavored.  Future high-statistics
top-quark production runs will likely make a more definitive statement.
The forward-backward asymmetry in $t\overline{t}$ production 
induced by axigluons in this mass window is also discussed and 
found to be quite large and so could provide another constraint.
\end{abstract}

\end{titlepage}
\double_spacesp

Extensions of the standard model of QCD which enlarge  the color group
to $SU_3(L) \times SU_3(R)$ at high energies, so-called chiral
color models \cite{glashow}, all predict the existence of a massive,
color-octet gauge boson which couples to quarks with an axial vector
coupling and the same strong interaction strength as QCD.  Such a particle
would make dramatic changes in heavy quarkonium decays and limits from
$\Upsilon$ decays involving  real \cite{cuypers_1}, \cite{robinett_1} or
virtual \cite{robinett_2} axigluons have been used to restrict their
mass to $M_A \geq 25\,GeV$ where $M_A$ is the axigluon mass. A more
stringent lower bound on $M_A$ can be derived by consideration of its
possible contribution to the $R$ ratio in $e^{+}e^{-}$ collisions with
a firm bound of $M_A \geq 50\,GeV$ now accepted \cite{cuypers_2}.
 Bagger, Schmidt, and King \cite{bagger} noted that axigluon production
in hadronic collisions could give rise to observable resonance structures
in the di-jet cross-section and various groups have found the following
overlapping limits on axigluon masses 
(usually with some assumptions made about the axigluon decay width)
\begin{center}
\begin{tabular}{lll}
UA1 & Ref.~\cite{ua1}          & $150 < M_A < 310\,GeV$ \\
CDF & Ref.~\cite{cdf_old}      & $120 < M_A < 210\,GeV$ \\
CDF & Ref.~\cite{cdf_dijets_1} & $200 < M_A < 870\,GeV$ \\
CDF & Ref.~\cite{cdf_dijets_2} & $200 < M_A < 980\,GeV$ \\
\end{tabular}
\end{center}
We combine these limits in Fig.~1 as an excluded band ($120< M_A<980\,GeV$) 
along  with the lower bound
from $e^{+}e^{-}$ interactions.  These bounds are the most restrictive ones
quoted in the most recent Review of Particle Properties \cite{pdg} and
indicate that there is still an allowed window for intermediate mass
axigluons in the range $50\,GeV < M_A < 120 \,GeV$.

Axigluons can also contribute to heavy flavor production by their
contribution to the sub-process $q\overline{q} \rightarrow g,A
\rightarrow Q\overline{Q}$ and in this note we consider what limits, if
any, top quark production data can set on intermediate mass (or very heavy)
axigluons.  The tree-level cross-section for top-quark production
from a $q\overline{q}$ initial state including both $s$-channel
gluons and axigluons is given by \cite{sehgal}
\begin{equation}
\left(
\frac{d \sigma}{d\hat{t}}
\right)_{q\overline{q}}
=
\left(
\frac{d \sigma}{d\hat{t}}
\right)_{0}
\left[
1 + |r(\hat{s})|^2 + 
4 Re(r(\hat{s})) \frac{(\hat{t} - \hat{u}) \hat{s} \beta}{ (\hat{t}-\hat{u})^2
+ \hat{s}^2 \beta^2} 
\right]
\label{cross-section}
\end{equation}
where $\beta = \sqrt{1-4M_t^2/\hat{s}}$ and $M_t$ is the top quark
mass.  The standard gluon-induced result is given by 
\begin{equation}
\left(
\frac{d \sigma}{d \hat{t}}
\right)_{0}
=
\frac{1}{16 \pi \hat{s}^2}
\frac{64\pi^2}{9} \alpha_s^2
\left[
\frac{(M_t^2-\hat{t})^2 + (M_t^2-\hat{u})^2 + 2M_t^2\hat{s}}{\hat{s}^2}
\right]
\end{equation}
and the enhancement factors due to axigluons depend on the ratio
\begin{equation}
r(\hat{s}) = \frac{\hat{s}}{\hat{s} - M_A^2 + iM_A\Gamma_A}
\end{equation}
where $\Gamma_A$ is the axigluon decay width.  We also 
have assumed that the axigluon-quark and gluon-quark  couplings 
both have strength $g_s$. The last term in square brackets in
Eqn.~(\ref{cross-section}) is due to the interference between the
gluon- and axigluon-induced processes and gives no contribution in
the total cross-section when integrated over parton distributions and
phase space as it is odd in $\cos(\Theta)$ where $\Theta$ is the
center-of-mass angle. 
  It does, however, contribute to a forward-backward asymmetry
in $t\overline{t}$ production \cite{sehgal} which we discuss below.  Any
increase in the top-quark production cross-section is then due to the
second term, $|r(\hat{s})|^2$, in Eqn.~(\ref{cross-section}).

In order to compare to
existing top-quark production data from the TEVATRON, we evaluate
the increase in the $t\overline{t}$ production cross-section due to this
term for various values of $M_A$; we use fixed values of $M_t = 175\,GeV$
and $\sqrt{s} = 1.8\,TeV$. Because top quark production
is dominated by the $q\overline{q}$ fusion mechanism at Born level, 
changes in this process can have a more obvious and direct 
effect on heavy quark production in this case than say 
for $b$-quark production where $gg \rightarrow b \overline{b}$ dominates. 
In fact, at the TEVATRON where the average center-of-mass energy for
$t\overline{t}$ production is much larger than $M_A$ in the intermediate
mass regime (namely $\langle \sqrt{\hat{s}} \rangle \approx 450\,GeV >
2M_t \approx 350\,GeV >> (50-120\,GeV)$, the effect 
of axigluons in this mass range would be
to more than double the already dominant tree-level $q\overline{q}$ 
production cross-section for top quark production 
since $|r(\hat{s})| \geq 1$ in this region.

The increase in the $t\overline{t}$
cross-section due to axigluons is plotted in Fig.~1 as a function of
$M_A$ for several different sets of assumptions; we normally
assume an axigluon decay 
width given by $\Gamma_A = N \alpha_s M_A/6 \approx 0.1M_A$ 
assuming $N=5$ light flavors of quarks as the main decay mode.  
The solid curve makes use of a set of parton distributions which is
extracted from data using a leading-order (LO) analysis 
\cite{owens} as might be considered most appropriate for this leading
order calculation. The dashed curve makes use of a recent, 
next-to-leading (NLO) set (MRRS, mode=1, $m_c\!=\!1.35\;GeV$, \cite{mrs}) 
and gives somewhat lower values; 
another popular NLO parameterization (CTEQ4M, \cite{cteq})
gives  almost identical results to this one.  The dot-dash curve uses
the LO parton distributions,  but assumes an axigluon width of
$\Gamma_A = 0.2M_A$; increasing $\Gamma_A$ has a negligible effect 
for small axigluon masses, but decreases the cross-section somewhat for
larger values of $M_A$. As mentioned above, for very low mass axigluons 
($M_A << 2M_t$), the cross-section
is essentially doubled from its $q\overline{q}$-induced tree-level value
and it is this region on which we will focus our attention. (The 
unacceptably huge resonant production contribution for $M_A \approx 2M_t$ 
would easily eliminate the region $200\,GeV < M_A < 780\,GeV$ which is
already excluded by di-jet analyses.) 
In the intermediate-mass range which is not already
excluded, the tree-level $q\overline{q}$-induced cross-section is
more than doubled and $\sigma_{t\overline{t}}$ 
would increase by $3.3-3.7\,pb$ ($2.7-3.1\,pb$) using LO (NLO)
parton distributions if axigluons in this mass range were present.

Starting with the first observation of the top quark at the TEVATRON
\cite{cdf_top}, \cite{d0_top}, there has been an increasingly large
set of data on the $t\overline{t}$ production cross-section in a 
variety of channels.  The experimental situation has been recently
reviewed by Gerdes \cite{gerdes_compare} who cites combined results 
(for dilepton and $l$+jets channels) for the D0 and CDF groups as
\begin{center}
\begin{tabular}{ll}
$\sigma_{t\overline{t}}(M_t = 173.3\,GeV) = 5.5 \pm 1.8\;pb$ &
D0 \\
$\sigma_{t\overline{t}}(M_t = 175.0\,GeV) = 7.5^{+1.9}_{-1.6} \; pb$ &
CDF \\
\end{tabular}
\end{center}
Increasingly more sophisticated QCD calculations of the top-quark
production cross-section, including resummation effects, have become
available and two recent groups find
\begin{center}
\begin{tabular}{lll}
Berger and Contopanagos & Ref.~\cite{berger} &
$\sigma_{t\overline{t}}(M_t=175\,GeV) = 5.52^{+0.07}_{-0.42}\;pb$ \\
Catani {\it et al} & Ref.~\cite{catani} &
$\sigma_{t\overline{t}}(M_t=175\,GeV) = 4.75^{+0.73}_{-0.62}\;pb$ \\
\end{tabular}
\end{center}
Given the relatively large statistical errors on the experimental
measurements, the theoretical and experimental values agree reasonably
well and arguably leave little room for a large new top-quark
production mechanism.   While we will not perform a detailed statistical
analysis, we can make some simple comments. 
If we average the experimental values and theoretical
predictions (combining their errors) and ask for the difference between
experiment and theory,  we find very roughly that
\begin{equation}
\overline{\Delta \sigma} 
=
\overline{\sigma_{t\overline{t}}(exp)}
-
\overline{\sigma_{t\overline{t}}(th)}
\approx 1.4 \pm 1.4\; pb
\end{equation}
which is certainly consistent with zero. 
This implies that an additional 
$\Delta \sigma_{t\overline{t}} = 3.2-3.7\,pb$ ($2.4-3.1\,pb$)
contribution from axigluon
intermediate states is disfavored, but only at the level of something
like $1.3-1.6\,\sigma$ ($0.9-1.2\sigma$) using LO (NLO) parton 
distributions in this leading-order analysis.
  The 50-fold increase in statistics expected for 
$t\overline{t}$ production during the next TEVATRON run (due to a
higher $\sqrt{s} = 2\,TeV$ and more luminosity) will dramatically reduce
the statistical errors quoted above and allow for a much better
comparison with both standard QCD and extensions thereof.  If the complete
set of NLO predictions for the combined gluon plus axigluon processes gave
rise to a similarly large increase in cross-section as has been observed
for the pure QCD case (something like a $44\%$ increase over LO), the 
disagreement with data  would increase even further.

It was noted some time ago \cite{sehgal} that axigluon intermediate
states could give rise to a large forward-backward asymmetry in heavy
quark production, due to the interference term in Eqn.~(\ref{cross-section}).
If we define $y = \cos(\Theta)$ as the center-of-mass angle for the
$2 \rightarrow 2$ $q\overline{q} \rightarrow Q \overline{Q}$ process, then
the forward-backward asymmetry (at Born level) can be defined as
\begin{equation}
A_{FB} = 
\frac{ 
\sigma_{q\overline{q}}(y>0) - \sigma_{q\overline{q}}(y<0)
}{
\sigma_{q\overline{q}}(y>0) + \sigma_{q\overline{q}}(y<0) + \sigma_{gg}
}
\label{asymmetry}
\end{equation}
where Born-level $gg$ process is evaluated using standard expressions
\cite{combridge}.  Once again, for fixed values of $M_t = 175\,GeV$, 
$\sqrt{s} = 1.8\,TeV$, LO parton distributions,  and $\Gamma_A = 0.1M_A$
 we plot this asymmetry in Fig.~2 versus 
$M_A$.  The forward-backward asymmetry in the allowed 
intermediate mass region of interest is very large ($\sim 75\%$) 
and does not vary much with parton distributions or axigluon width.
More detailed tests of the top quark production process might 
well be used to probe this effect and further constrain axigluon masses.

\begin{center}
{\Large
{\bf Acknowledgments}}
\end{center}

One of us (M.A.D) acknowledge the support of Penn State University 
through a Research Development Grant (RDG).

\newpage
{\Large
{\bf Figure Captions}}
\begin{itemize}
\item[Fig.\thinspace 1.] Additional contribution to the $t\overline{t}$ 
production cross-section due to axigluons (in $pb$) for $p\overline{p}$
collisions at $\sqrt{s} = 1.8\,TeV$ using $M_t = 175\,GeV$ versus mass
of the axigluon ($M_A$) in GeV.  The curves correspond to 
LO parton distributions \cite{owens} and $\Gamma_A = 0.1M_A$ (solid),
NLO partons \cite{mrs} and $\Gamma_A = 0.1M_A$ (dashed), and
LO partons \cite{owens} and $\Gamma_A = 0.2M_A$ (dot-dash).
Excluded regions for the axigluon
mass from collider dijet data and $e^{+}e^{-}$ analyses are also shown.
\item[Fig.\thinspace 2.] Forward-backward asymmetry $A_{FB}$, defined in
Eqn.~(\ref{asymmetry}),  for $t\overline{t}$
production in $p\overline{p}$ collisions (assuming $M_t = 175\,GeV$
and $\sqrt{s} = 1.8\,TeV$) versus axigluon mass using LO partons
\cite{owens} and $\Gamma_A = 0.1M_A$. Also shown are
published limits on the axigluon mass.
\end{itemize}


\newpage
\begin{thebibliography}{99}
%
\bibitem{glashow} P. H. Frampton and S. L. Glashow, Phys. Lett.
{\bf 190B}, (1987) 157; Phys. Rev. Lett. {\bf 58}, (1987) 2168.
%
\bibitem{cuypers_1} F. Cuypers and P. H. Frampton, Phys. Rev. Lett.
{\bf 60}, (1988) 1237.
%
\bibitem{robinett_1} M. A. Doncheski, H. Grotch, and R. Robinett,
Phys. Lett. {\bf 206B}, (1988) 137.
%
\bibitem{robinett_2} M. A. Doncheski, H. Grotch, and R. W. Robinett,
Phys. Rev. {\bf D38}, (1988) 412.
%
\bibitem{cuypers_2} F. Cuypers, A. F. Falk, and P. H. Frampton,
Phys. Lett. {\bf 259}, (1991) 173. 
%
\bibitem{bagger} J. Bagger, C. Schmidt, and S. King, Phys. Rev.
{\bf D37}, (1988) 1188.
%
\bibitem{ua1} C. Albajar {\it et al.} (UA1 Collaboration), Phys. Lett.
{\bf 209B}, (1988) 127.
%
\bibitem{cdf_old} F. Abe {\it et al.} (CDF Collaboration), Phys. Rev.
{\bf D41}, (1990) 1722.
%
\bibitem{cdf_dijets_1} F. Abe {\it et al.} (CDF Collaboration),
Phys. Rev. Lett. {\bf 74}, (1995) 3538.
%
\bibitem{cdf_dijets_2} F. Abe {\it et al.} (CDF Collaboration) 
Phys. Rev. {\bf D55}, (1997) R5263.
%
\bibitem{pdg} Review of Particle Physics, Particle Data Group,
Phys. Rev. {\bf 54}, (1996) 1; see page 236 for the detailed 
review. 
%
\bibitem{sehgal} L. M. Sehgal and M. Wanninger, Phys. Lett. {\bf 202B},
(1988) 211.
%
\bibitem{owens} J. F. Owens, Phys. Lett. {\bf B266}, (1991) 126. 
%
\bibitem{mrs} A. D. Martin, R. G. Roberts, M. G. Ryskin, and W. J. Stirling,
Durham preprint DTP-96-102, hep-ph/9612449.
%
\bibitem{cteq} H.L. Lai, J. Huston, S. Kuhlmann, F. Olness, J. Owens, 
D. Soper, W.K. Tung, and H. Weerts, Phys. Rev. {\bf D55}, (1997), 1280.
%
\bibitem{cdf_top} F. Abe {\it et al}, Phys. Rev. Lett. {\bf 74}, (1995)
2626.
%
\bibitem{d0_top} S. Abachi {\it et al.}, Phys. Rev. Lett. {\bf 74},
(1995) 2632.
%
\bibitem{gerdes_compare} D. W. Gerdes (for the CDF collaboration), 
FERMILAB-CONF-97/166-E, hep-ex/9706001, 
to appear in the proceedings of the XXXIInd 
Rencontres de Moriond, Electroweak Interactions and Unified Theories, 
Les Arcs, Savoie, France, March 1997.
%
\bibitem{berger} E. L. Berger and H. Contopanagos, Phys. Rev. {\bf D54},
(1996) 3085; Argonne report ANL-HEP-CP-97-33, hep-ph/9706356. 
to appear in the Proceedings of DIS97, Fifth International Workshop 
on Deep Inelastic Scattering and QCD, Chicago, IL, April 1997. 
%
\bibitem{catani} S. Catani, M. L. Mangano, P. Nason, and L. Trentadue,
Phys. Lett. {\bf B378}, (1996) 329.
%
\bibitem{combridge} B. Combridge, Nucl. Phys. {\bf B151}, (1979) 429.
%
\end{thebibliography}
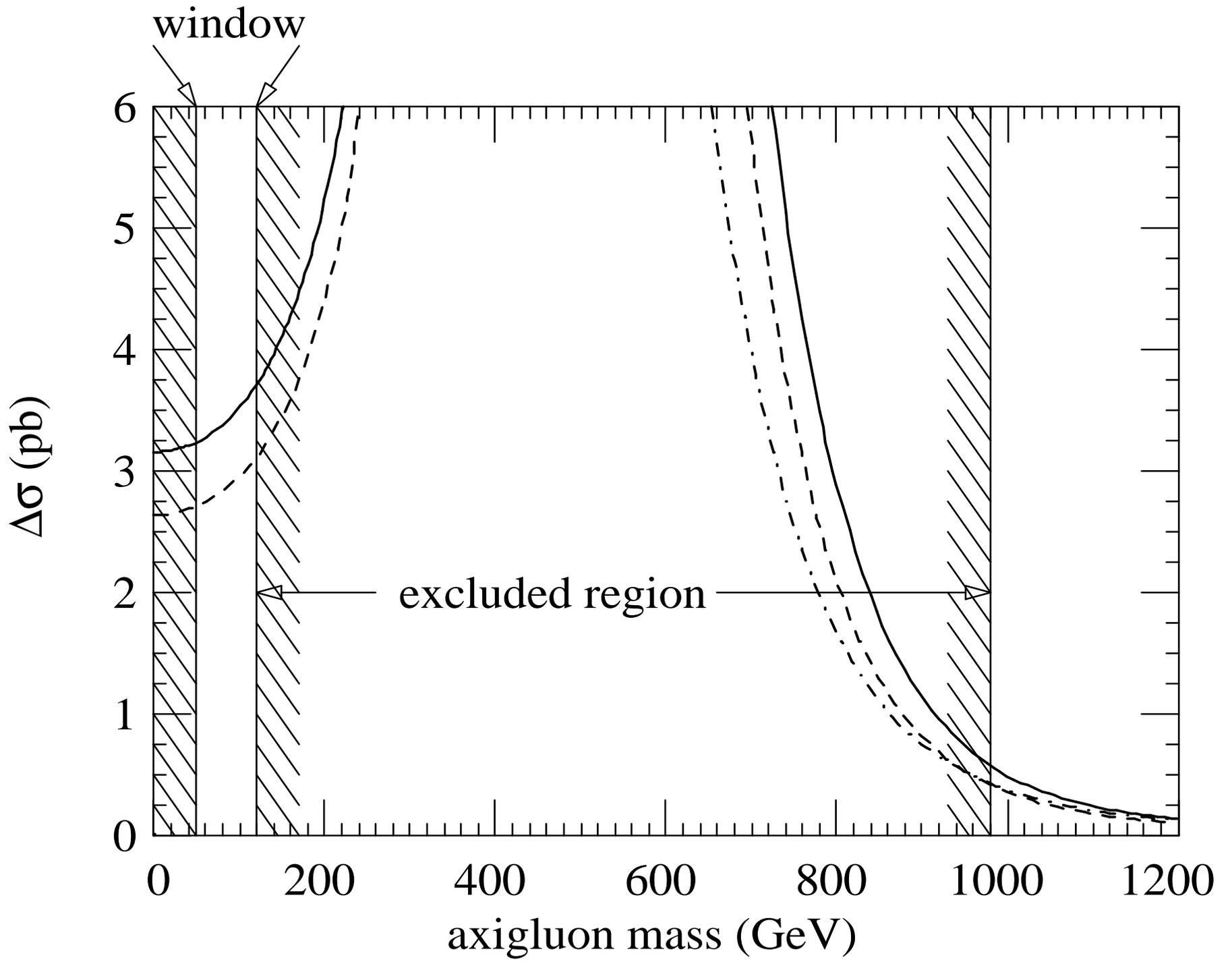
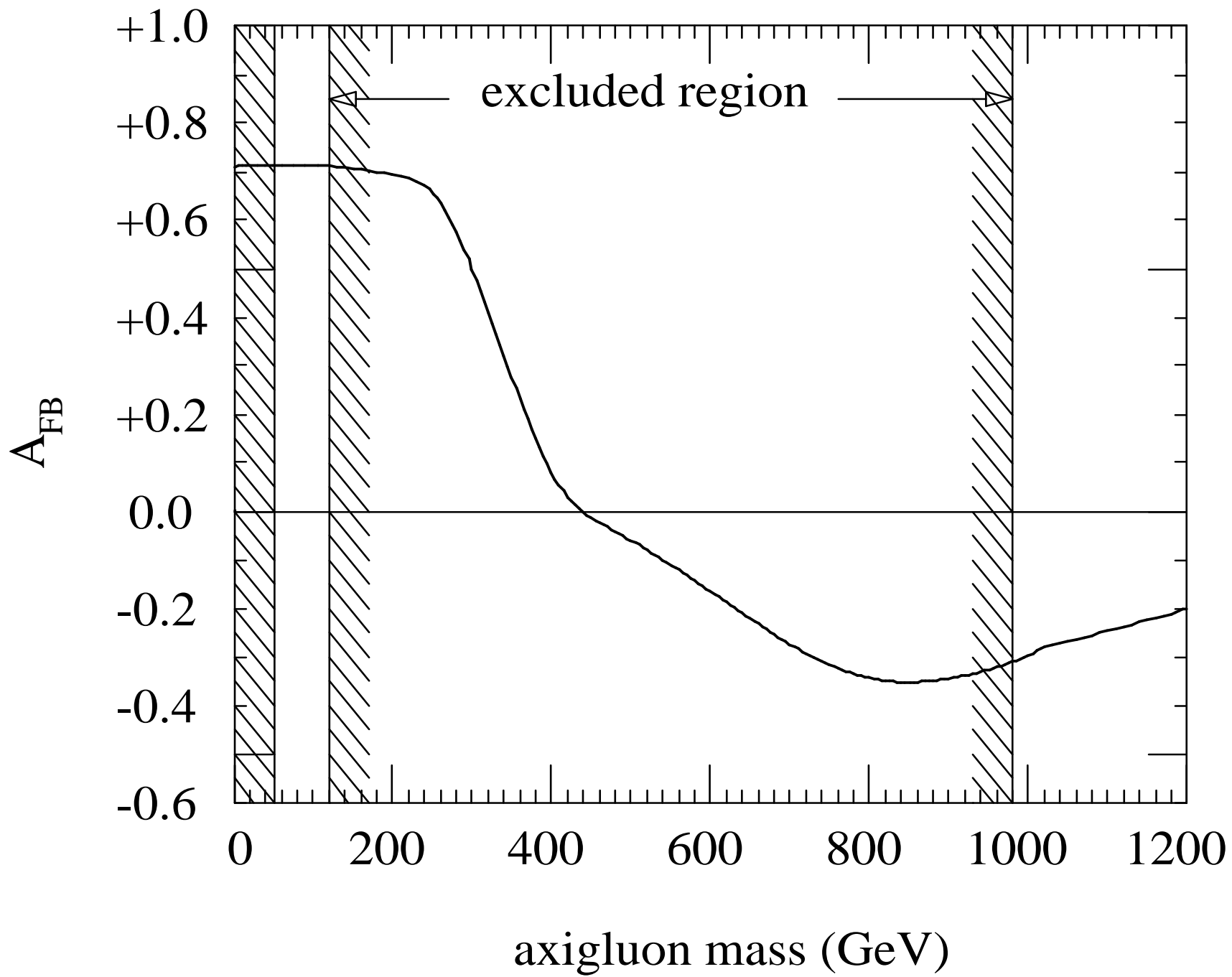
\end{document}